\documentclass[10.5pt]{article}

\usepackage{mathptmx}
\usepackage{graphicx}
\usepackage{times}
\usepackage{amsmath, amsthm, amssymb}
\usepackage{url}
\usepackage{subfig}
\usepackage{hyperref}
\usepackage{algorithm}
\usepackage{algorithmic}

\usepackage{lscape}

\begin{document}

\begin{titlepage}

\begin{center}
{ \Large \bf
{The Dynamics of Emotional Chats with Bots: Experiment and Agent-Based Simulations}
}

{\large Bosiljka Tadi\'c$^1$, Vladimir Gligorijevi\'c$^{1}$, Marcin Skowron$^3$ and Milovan \v Suvakov$^{1,2}$}\\
{ $^1$Department of theoretical physics; Jo\v zef Stefan Institute;
 Box 3000  SI-1001 Ljubljana Slovenia; $^2$Institute of Physics, Belgrade, Serbia, $^3$Austrian Research Institute for Artificial Intelligence, Freyung 6/6, 1010 Vienna, Austria\hspace{1cm}}

\date{today}

\end{center}

\noindent
\begin{abstract}Quantitative research of emotions in psychology and machine-learning methods for extracting emotion components from text messages open an avenue for physical science to explore the nature of stochastic processes in which emotions play a role, e.g., in human dynamics online. Here, we investigate the occurrence of collective behavior of users that is induced by chats with emotional Bots. The Bots, designed in an experimental environment, are considered. Furthermore, using the agent-based modeling approach, the activity of these experimental Bots is simulated within a social network of interacting emotional agents. Quantitative analysis of time series carrying  emotional messages by  agents suggests temporal correlations and persistent fluctuations with clustering according to emotion similarity.
{\bf {All data used in this study are fully anonymized.}}
\end{abstract}
\vspace{0.5cm}
\tableofcontents
\end{titlepage}
\newpage
\section{Introduction\label{sec-intro}}
Recently, quantitative research of human dynamics on Web has been enabled thanks to a large amount of empirical data that are systematically accumulated at the various Web portals. This opened a new avenue of quantitative social science \cite{nature2011}. Online social dynamics constitute a complex system, in which micro-conceptions of global behaviors only began to reveal the working mechanisms of  new emergent phenomena \cite{ST_Plos2013,we-Entropy}. In analogy to complex systems in the physics laboratory, research methods and concepts of statistical physics are essential for finding how the global features emerge from particular actions in human interacting systems. However, in contrast to the laboratory physical systems, human qualities---i.e., cognition and emotion---play an important role in forming the dynamics  at a microscopic scale. In order  to take into account these features of human dynamics correctly,  in recent years interdisciplinary research approaches are being developed. Specifically, the methods of statistical physics and graph theory, on one side, in synergy with machine-learning methods of text analysis, on the other, have been considered. It should be stressed that, each of these approaches alone takes into account a specific aspect of the human dynamics and potentially avoids the whole story. In particular, machine-learning methods can extract information and emotion from written text (for a recent review of methods see \cite{Liu2012}); however, this feature, observed at microscale (i.e., in each message separately),  need to be suitably incorporated into a stochastic process in order to see its effects at a large scale. Similarly, the physics theory of complex systems, without these microscopic features of individual users and their messages over time, would remain at phenomenological level, thus without a predictive power for a concrete social system. Moreover, the graph-theory methods, which provide a formal mathematical framework for quantitative analysis of these systems, need a thoughtful concept from the theory of complex systems in order to, firstly, build an appropriate graph of the considered dynamical system and, secondly, to concentrate on its relevant features.

These three methodologies used in synergy have been successfully applied in the quantitative analysis of collective social behaviors in which emotions play a role, specifically in the empirical data from  Blogs \cite{mitrovic2010a,dodds2011}, Diggs \cite{mitrovic2011}, Forums \cite{warsaw2011}, Online games \cite{Szell2010_,szell2010}, Online social networks \cite{we-MySpace11,myspace2007,facebook2011}, \texttt{Twitter} \cite{twitter2011}, Online chats \cite{we-Chats1s,garas2012} and others. It has been shown that, depending on the type of interaction between users (direct versus indirect) in accordance with a specific use of emotions which are expressed in the text messages that are exchanged between users, different classes of online social networks emerge through the self-organized dynamics on Web portals \cite{grujic2009,mitrovic2010a,mitrovic2010b,we-MySpace11,Szell2010_,szell2010,mitrovic2010c,we-Chats-conference,we-Chats-chapter}.  Apart from online social networks \texttt{MySpace} and \texttt{Facebook} where  a priori a kind of social graph exists, in all other circumstances  the co-evolution of a network structure occurs together with the underlying stochastic process of human interactions \cite{we-Entropy}.

In analogy to the above mentioned data analysis, in mathematical modeling of social dynamics a list of human attributes needs to be taken into account  \cite{BT_ABMbook}. For these purposes, the appropriate framework is provided by the agent-based modeling approach, where emotional agents are located on (evolving) social networks \cite{BT_ABMbook,we_ABMblogs12,we-MySpaceABM,we-ABMrobots,we-Entropy}. In addition to human users, the presence of artificial systems, e.g. Bots is common in the virtual world. The Bots with different functions (suggesting goods, trading on the stock market, social 'gaming', etc) may interact with the real users either directly or indirectly. In these interactions, often requiring an emotional impact, the Bots may become very effective, outperforming human users. The underlying mechanisms that help the Bots' performance  rely on the long-range correlations among users and  efficient propagation of emotion through the social graph \cite{we-ABMrobots}.

In this work, we employ recently developed agent-based model of  chats with emotional Bots in order to \textit{ extend} an experiment with Bot--user dialogs 
into the interacting environment of social networks. The experiment that we consider was done with three emotional Bots, i.e., Affective Dialog Systems \cite{ACII1} and a limited number of separated users in their natural environments.  In the remaining part of the paper, we first describe the experiment and use graph-theory methods to reveal the structure in the experimental data. Then, we implement the same experimental Bots into the system of interacting agents; performing the simulations and analyzing the simulated data,  we get a number of new results that quantify the Bot's impact on the entire network of agents. Finally, to evaluate the network effect onto  the agents which are in direct contacts with the Bot, we compare the simulated data of these interacting agents with the empirical data containing the same number of isolated users.

\section{Chats with emotional  Bots in a controlled environment\label{sec-patternExp}}
The parameters used in the simulation (see section\ \ref{sec-abmBotExp}) are extracted from experiments 
in which users interacted with a variant of \textit{conversational Bot - Affective Dialog System} \cite{Skowron2010}. 
The aim of the experiment was to study how different affective profiles i.e., positive, negative, neutral \cite{ACII1}
used by the Bot influence communication patterns, affective responses of participants and further how do they relate to the experienced emotional changes.

Below we present and overview of the system applied in the experiment and closer introduce the concept of {\it artificial affective profile}.
The detailed description of the system software framework and its core components is presented in \cite{FLAIRS,TAFFC2013};
the extensive overview of settings and results obtained in this round of experimens in \cite{ACII1,ACII2}.
\subsection{Experiment with Affective Dialog System\label{sec-experiment}}
The Affective Dialog System (DS) communicates with users in a textual modality, in a natural language form, and
uses integrated affective components for detecting textual expressions of users' affective states.
The system interacts with users via a range of communication channels and interfaces
that share common characteristics of online chatting.

At the top-level the system architecture includes 3 main layers: perception, control and communication. 
The perception layer annotates user's and system's response candidates. 
It uses a set of natural language processing and affective processing tools and resources \cite{WACI}.
Based on the information cues from the perception layer, the control layer selects and, if required, alters system response candidates.
Also, the layer manages dialog progression accounting for its context and the applied affective profile. 
The control layer uses an information state based dialog management component: Affect Listener Dialog Scripting (ALDS) \cite{Skowron2010} for the closed-domain and task-oriented parts of the dialog. 
For the open-domain chats, a template based mechanism and response generation instructions - AB-AIML \cite{WACI}, are applied. 
The system's affective profile influences the selection of both ALDS scenarios and subsets of AB-AIML response instructions. 
For the remaining system response candidates for which no specific affective profile dependent interaction scenarios or response instructions are available,
a post-precessing is applied, i.e., addition, removal of positive or negative words.
This mechanism aims at aligning the affective load, i.e. valence of system response candidates with the selected affective profile \cite{ACII1}.

For the purpose of the experiment, an \emph{artificial affective profile}  was defined as a coarse-grained emulation
 of affective characteristics of an individual,
corresponding to dominant, observable affective traits, that can be consistently conveyed
by a conversational Bot during the course of its interactions with users \cite{ACII1}.
In particular, three distinct affective profiles were implemented --- labeled as positive, negative and neutral.
Each affective profile aimed at a consistent demonstration of character traits of the system
that are described as, respectively:
\begin{itemize}
\item{\texttt{positiveBot}}: {polite, cooperative, empathic, supporting, focusing on similarities with a user;}
\item{\texttt{negativeBot}}: {conflicting, confronting, focusing on differences with a user;}
\item{\texttt{neutralBot}}: {professional, focused on the job, not responding to expressions of affect.}
\end{itemize}
The study directly addressed the research questions about the artificial system's ability
to consistently imitate affective profiles, validates the methods applied and further
enables to measure their effects in interactions \cite{TAFFC2013,SkowronRankInPress}.

\subsection{Experimental Settings and Results}

During the experiment, participants were in their ``natural environment'' - a place where they usually use the Internet, 
which is assumed to make them more receptive as well as spontaneous.
For the majority of participants, English, the language in which the experiments were conducted, was not their native language,
but all users who completed the set of interactions had at least average communicative skills in this language.
The usage of non-native languages in online interaction environments is a frequent phenomenon 
and provides the motivation for studying this type of communication.

The study was conducted using a browser-based communication interface,
resembling a typical web chat-room environment:
a user input field at the bottom of the screen and a log of communication in the center.
Participants interacted with the Bot which applied three different affective profiles (positive, neutral and negative)
in turn, once with each.
To avoid ordering effects in the evaluation of the different system realizations, the actual sequence was randomly
and evenly assigned.
The evaluation statements were also displayed to users before the start of the first interaction to familiarize themselves with rating.
These statements were:
\begin{enumerate}
\item{I enjoyed chatting with the conversational partner during the just completed interaction.}
\item{I found a kind of "emotional connection" between myself and the conversational partner.}
\item{I found the dialog with the conversational partner to be realistic.}
\item{I found the dialog to be coherent. In other words, the sequence of responses of the conversational partner made sense.}
\item{I noticed a positive emotional change in myself during the interaction.}
\item{I noticed a negative emotional change in myself during the interaction.}
\item{I would like to chat again with this particular conversational partner in the future.}
\end{enumerate}

After each experimental condition i.e., interaction with the Bot which applied a selected affective profile,
participants were asked to express their agreement or disagreement with the statements
on a five-point Likert scale, i.e.,
from 1 $=$ \emph{strongly disagree} to 5 $=$ \emph{strongly agree}.
Participants were aware that they were chatting with an artificial system.
They interacted with the Bot in an unsupervised manner.
Interactions were always initiated by the Bot, i.e. the system provided the first utterance,
and stopped after 7 minutes with a closing response,
followed by the display of the questionnaire.

Before the experiment was initiated, the participants were instructed to freely chat with the Bot.
No additional, more detailed guidelines were given to avoid artificial constraints, e.g.,  
the communication strategy applied by participants.
Interactions were completed by 91 participants (33 female, 58 male), aged between 18 and 52, in all three experimental settings resulting in 273 interaction logs.

\textit{Data annotation:} The analysis of the presented data-set was conducted with a set of natural language processing and affective processing tools and resources,  
including: Support Vector Machine Based Dialog Act classifier\cite{ACII2}, Lexicon Based Sentiment Classifier\cite{gpalt}, 
Linguistic Inquiry and Word Count dictionary\cite{LIWC},  ANEW dictionary based classifier \cite{ANEW}. 
Further, we analyzed timing information and surface features of participants communication style such as wordiness and usage of emoticons. 

For the purpose of this work, we use the emotional attributes---arousal and valence---of each message in the dialogs with Bots. 
These attributes are determined  by the classifier which uses 
{\bf ANEW}\label{ANEW}---Affective Norms for English Words dictionary. 
The dictionary is based on the assumption that emotion can be defined as a coincidence of values on a number of strategic dimensions \cite{ANEW}
and includes a set of 1,034 commonly used words, including verbs, nouns and adjectives.
It provides information on emotional content of an input string, in three affective dimensions: valence, arousal and dominance, 
on the scale from 1 (very unpleasant, low arousal, low dominance/control) to 9 (very pleasant, high arousal, high dominance/control).
For example ``shark'' has the following mean score for the three presented affective dimensions (valence - 3.65, arousal - 7,16, dominance - 2.16).
The scores for ``victory'': (valence - 8.32, arousal - 6,63, dominance - 7.26).

\begin{figure}[htb]
    \centering
    \includegraphics*[scale=.45]{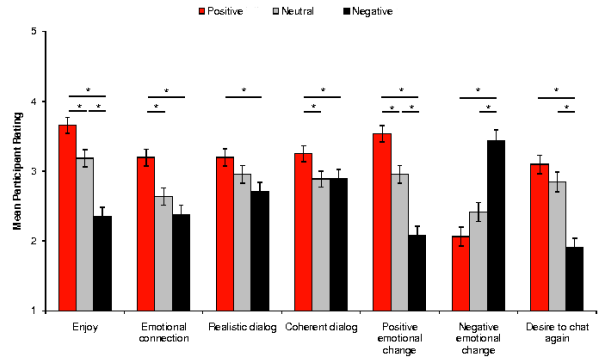}
    \caption{Participant's mean ratings on all dependent variables,
        of their interactions with the dialog system (DS) with three different
        affective profiles (positive, neutral, negative). [Data from \cite{ACII1}].
    }
    \label{stat2}
\end{figure}

\textit{Effects of Affective Profile on System's Evaluation and Emotional Changes:} The affective profile had a series of significant effects on the evaluation of the Bot
and on users' emotional changes.
Detailed results of the analyses performed are presented in \cite{ACII1} and \cite{ACII2}.
Concerning evaluation, the affective profile had significant effects on all dependent measures
(see Figure~\ref{stat2}).
The largest effect sizes were found on statements 5 and 6
(positive and negative emotional change, respectively).
As expected, the Bot's affective profiles successfully induced corresponding emotional changes in users,
affecting perception of dialog realism and coherence only to a smaller extent.

\textit{Effects of Affective Profile on Users' Interaction Style and Expressions of Affective States:}
Users were equally fast in replying to different affective profiles.
In particular, when analyzing the whole interactions, there were no differences in the participants' average response time to a number of letters generated.
Participants also used an equal amount of words and utterances in their conversations with the Bot, regardless of the applied affective profile.
There were, however, significant differences in word categories used and other linguistic aspects of the text input. 

\begin{figure}[htb]
    \centering
    \includegraphics*[scale=.45, trim=0 0 0 0]{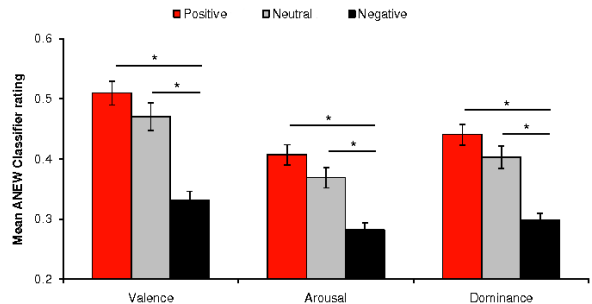}
    \caption{Mean valence, arousal, and dominance ratings obtained from ANEW classifier for user exchanges with the dialog system (DS).
    [Data from \cite{ACII2}].
    }
    \label{SCANEW}
\end{figure}

The analysis of ratings obtained from ANEW classifier for participants utterances, 
revealed the effect of the applied affective profile.
Specifically, the negative affective profile elicited lower valence, arousal and dominance, 
i.e. mean ratings from ANEW classifier assigned to the participants' utterances
(see Figure~\ref{SCANEW}).
Consequently, it can be seen, that users in their interactions with the Negative version of the Bot assumed rather passive stance towards it 
and neither attempted to retaliate nor dominate it.

Conclusion: Among others, compared with the positive profile, the negative profile elicited less assent (e.g., ok, yes, agree) from users, fewer positive emotion words,
more anger-related words, and in overall, significantly lower valence.
The positive profile, contrary, elicited accordingly more positive emoticons,
more positive emotion words (e.g., nice, love), more user statements, and less closed questions to the system.
The two latter findings also indicate more information disclosure and less questioning from users towards the positive profile, compared to the negative profile.
The three different affective profiles used by the Bot did not influence users' interaction style, in aspects related to e.g. response time or wordiness. 

\subsection{Network analysis reveals the structure of emotional chats \label{sec-empirical}}
\begin{figure}[ht]
\begin{tabular}{ccc}
 {\large (a)}& {\large (b)}\\
\resizebox{18pc}{!}{\includegraphics{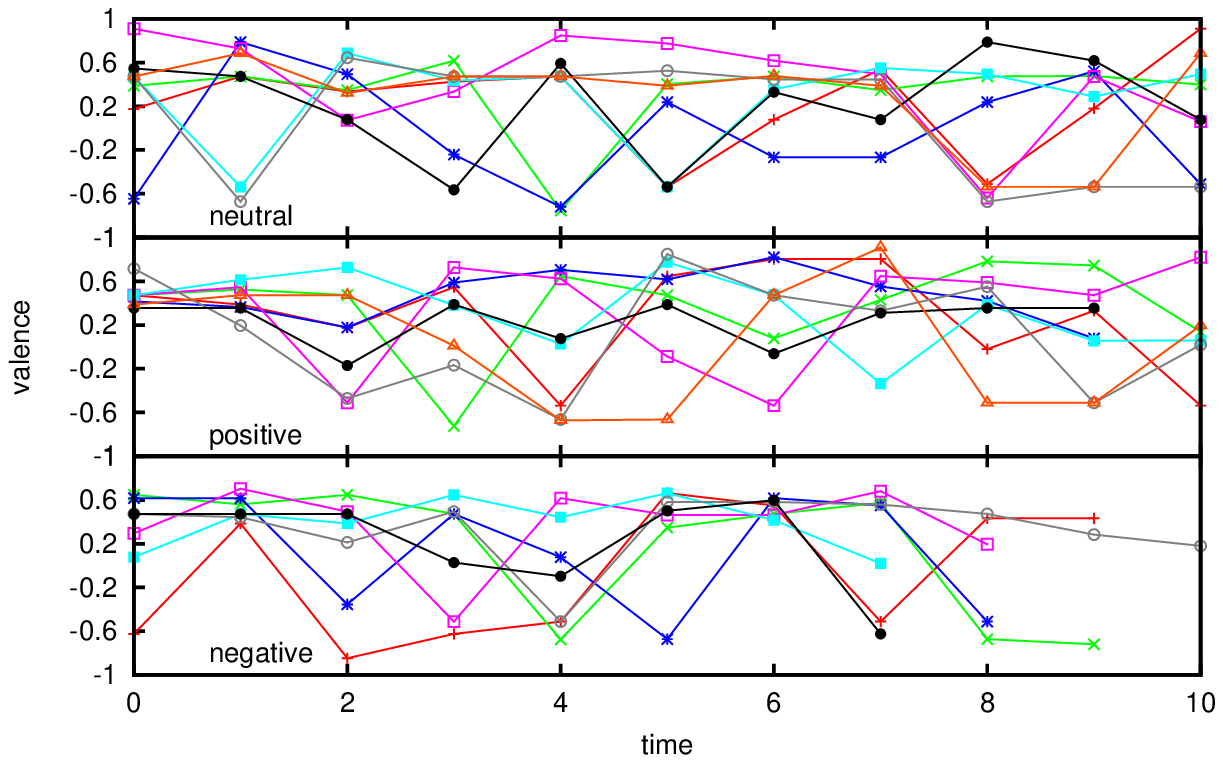}}&
\resizebox{16pc}{!}{\includegraphics{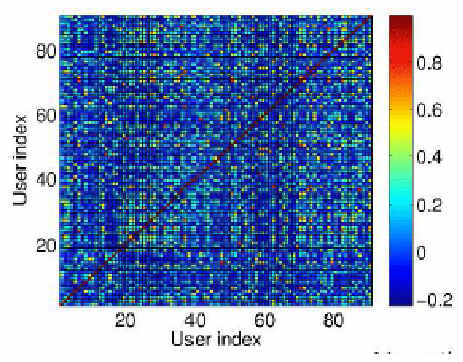}}\\
\end{tabular}
\caption{(a) Examples of time series of valence in the user's response to different Bots. Messages by the same user are indicated by the same color in all panels. (b) The filtered correlation matrix in the case of positive Bot. }
\label{fig-matrixExp}
\end{figure}

For each of three Bots, the empirical data consists of 91 pairs of time series of a Bot's messages (arousal and valence sequence) and the corresponding user's arousal and valence time series. For example, the fluctuations of emotional valence deduced from the answer of the first seven users in the dialogs with the  emotional Bots are shown in Fig.\ \ref{fig-matrixExp}a. In total, we have 3x91x2x2 separate sequences with emotional content to be analyzed. Here, we use network theory methods to extract patterns of similar response of users to each particular profile of the Bot. Creating a correlation matrix is a first step to infer such information from multichannel data; it has been practiced in the analysis of  large throughput gene expression experiments \cite{zivkovic2006,izraelci}, classifying similar force-distance curves in repeated force-spectroscopy measurements of molecular forces \cite{jelena-EPL,jelena2013}, finding relevant correlations in fluctuations of different shares in stock market \cite{burda} and identifying patterns in traffic jamming \cite{tadic2009,hongli2009}.
Here, we construct Pearson's coefficient $C_{ij}$ of the valence time series between all pairs of users, $i,j=1,2\cdots N$, $N=91$,  and focus on positive correlations only.

\begin{figure}[ht]
\begin{tabular}{ccc}
 {\large (a)}& {\large (b)}& {\large (c)}\\
\includegraphics[scale=0.52]{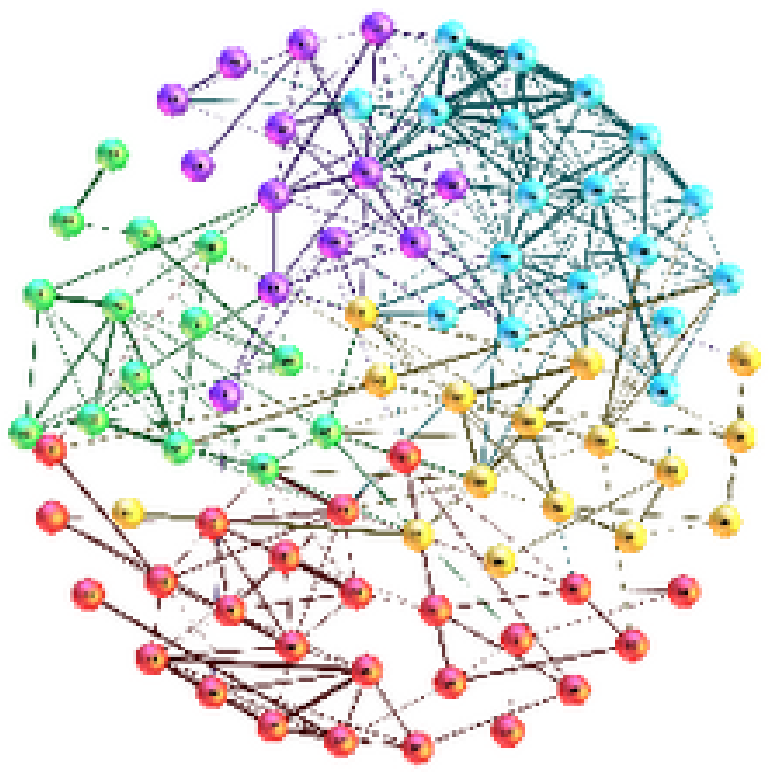}&
\includegraphics[scale=0.52]{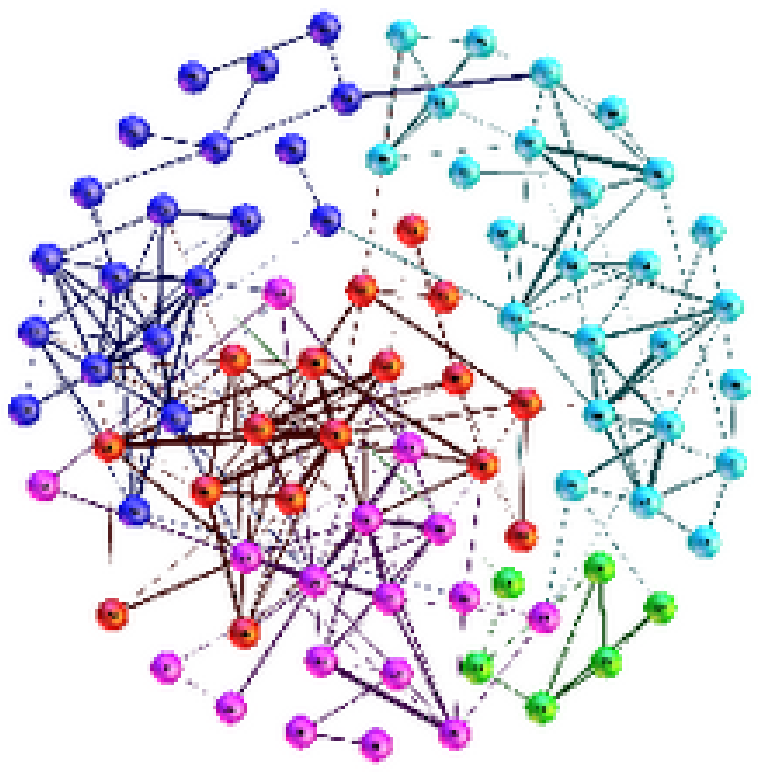}&
\includegraphics[scale=0.52]{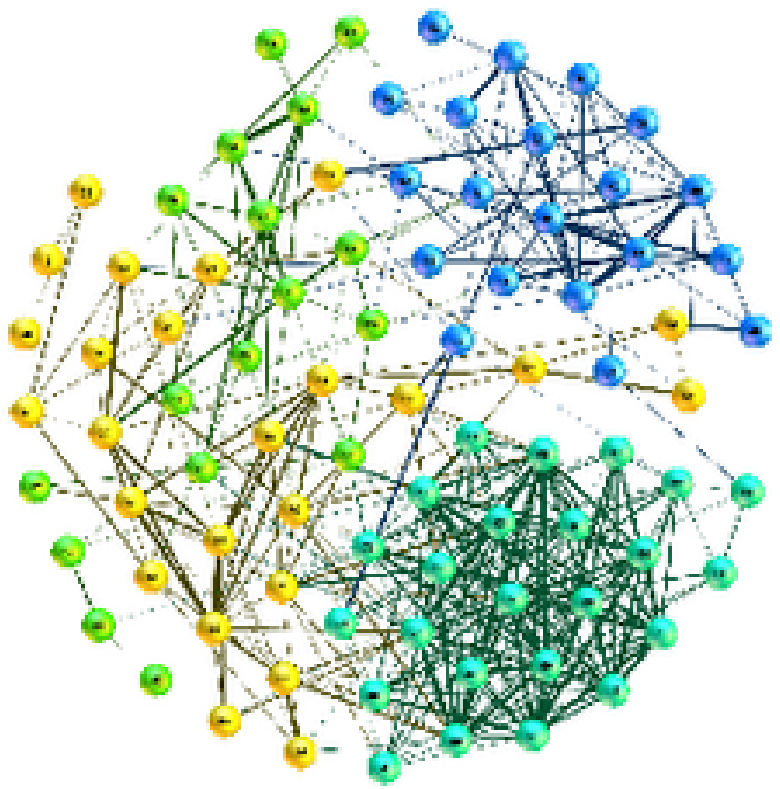}\\
\end{tabular}
\caption{Valence correlation networks of 91 users chatting with  a Bot with (a) positive, (b) negative   and (c) neutral emotional profile. Threshold: $C_0=0.4$. }
\label{fig-groupsExp}
\end{figure}

The time series of valence $\{v_i(t_k)\}$ for each user $i=1,2, \cdots 91$ are available in the experimental dataset. We construct the correlation matrix from these time series as
\begin{equation}
C_{ij}= \frac{\sum_k[v_i(t_k)-<v_i>][v_j(t_k)-<v_j>]}{\sigma_i\sigma_j} \ ;
\label{eq-correlation}
\end{equation}
where $<v_i>$ and $\sigma_i$ are the average and standard deviation of the valence fluctuations of the user $i$ during the entire time of the conversation with the Bot, $t_k=1,2,...t_e$.  
Following a standard procedure \cite{izraelci,tadic2009,jelena-springer,jelena2013}, the correlation matrix is further filtered to reduce spurious correlations and enhance the correlations $V_{ij}$ among such pairs of users who correlated with the rest of the system in a similar way. Then a threshold $V_0$ is applied, and the correlations exceeding the threshold are considered.
The filtered correlation matrix with the elements  $C_{ij}>C_0$ over a noise threshold corresponding to the case of dialogs with the positive Bot is shown  in Fig.\ \ref{fig-matrixExp}b.
The correlation matrix can be also represented as a mathematical graph (network) with weighted symmetrical links. Three networks corresponding to the three experimental Bots are shown in Fig.\ \ref{fig-groupsExp}. Applying the community detection methods based on the maximum-modularity algorithm \cite{lancichinetti2010}, different communities of users are detected.
Different colors of nodes in Fig.\ \ref{fig-groupsExp} on each network indicates a \textit{community}---group of users who expressed similar fluctuations in the emotional valence as detected  in their response to the Bot's message.

It is interesting to note that, above a reasonable correlation  threshold, five different groups of users can be detected based on their similarity in the expressed emotions to the Bot with the same profile. Also, these groups may contain different users when the Bot's emotional profile is changed, for instance, from positive to negative. The occurrence of groups can be expected on the basis of different psychology profiles \cite{ryan2011} of the users, i.e., shyness, narcissism, loneliness etc, which may affect a user's emotion expression. 
  (Unfortunately, the profiles of users who participated in the experiment were not examined.)  This aspect of emotion expression may be even enhanced in the conditions of isolated user--Bot dialog as it was the case in the experiment.

\section{The emotional Bots in social chat-networks: Agent-based simulations
\label{sec-abmBotExp}}
\begin{figure}[ht]
\begin{tabular}{ccc}
\resizebox{18pc}{!}{\includegraphics{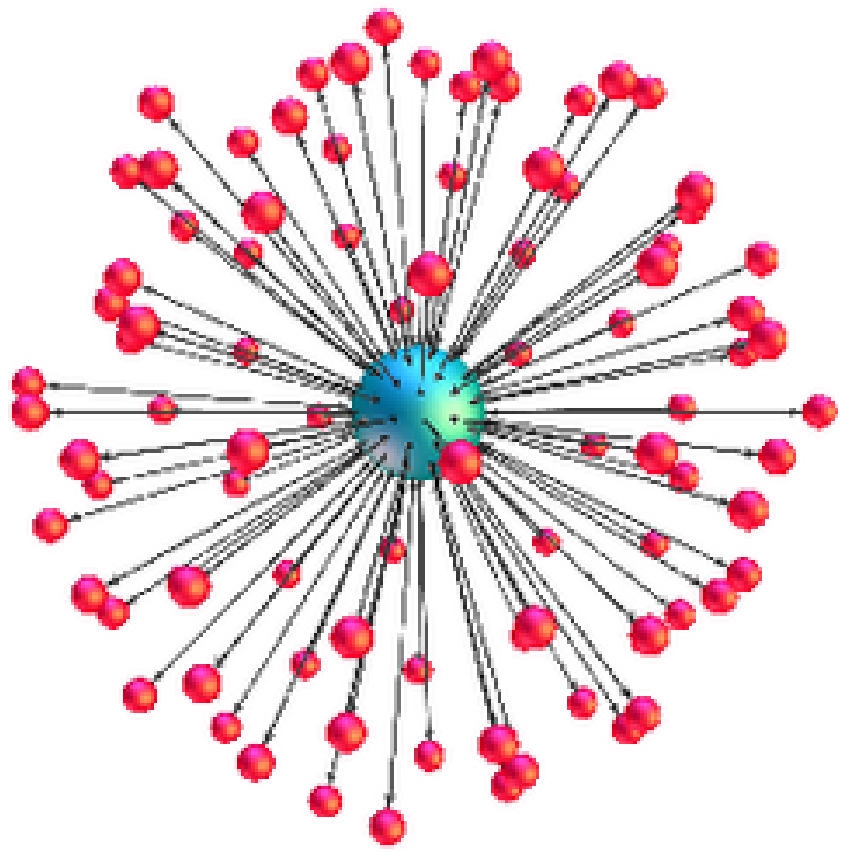}}&
\resizebox{18pc}{!}{\includegraphics{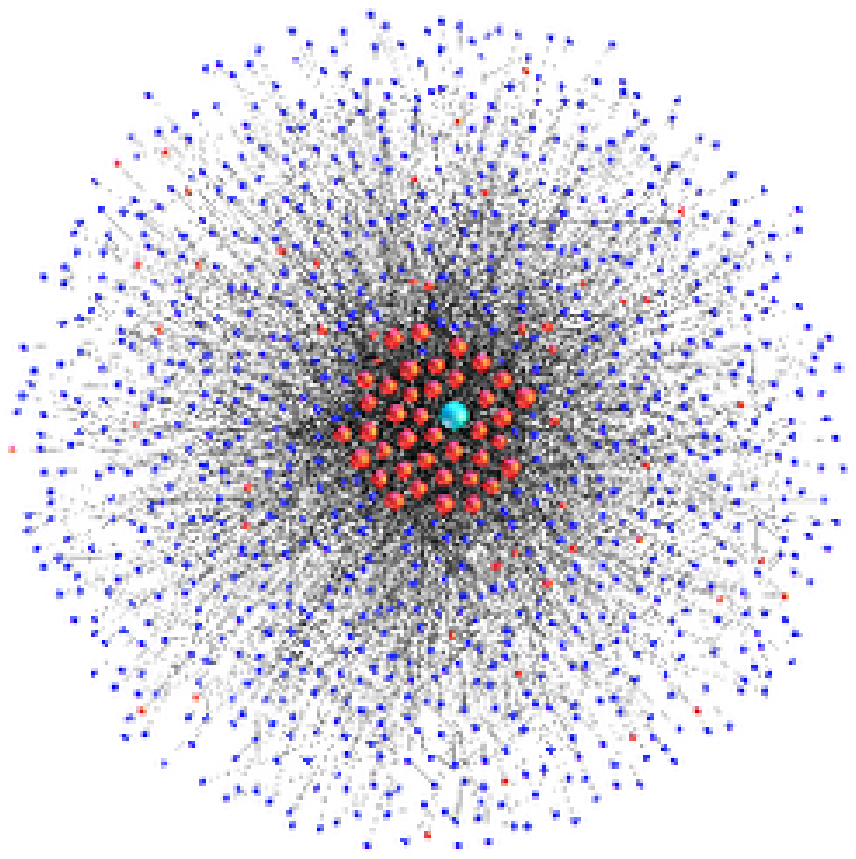}}\\
\end{tabular}
\caption{Left: Star network of 91 users individually chatting with the Bot (central node). Right: The experimental Bot (central node) and 91 agents receiving the Bot's emotional messages (red nodes) as a part of network of agents (blue nodes) simulated within the agent-based model (see sect.\ \ref{sec-abm} for details) }
\label{fig-network_exp_abm}
\end{figure}

In this section, we examine the effects of social environment using agent-based modeling we simulate the situation when such  users (represented by agents) are involved in the emotional chats with the same Bots but interact with other agents on an evolving social network. The different environments are represented by networks in Fig.\ \ref{fig-network_exp_abm}. On left, the star-like network with 91 users linked with the Bot (in the middle) represents mutually non-interacting  users in the controlled experiment as described in sect.\ \ref{sec-experiment}. The figure on the right pictures an emerging network of chats. The Bot (in the middle) and these 91 users,  which are designed by 91 agents (red colored nodes),  are situated in an interacting environment where they can chat with each other and with other agents in the network.

\subsection{Features of the agent-based model of emotional chats with Bots\label{sec-abm}}
To simulate the effects of these emotional Bots in a social network, here we use the agent-based model of chats between an emotional Bot and agents that we have developed recently \cite{we-ABM_bots2,we-ABMrobots}. For the purpose of this work, we first briefly describe the relevant features of the model. More detailed description, the dynamic rules as well as the numerical implementation of the basic model can be found  in \cite{we-ABMrobots,we-ABM_bots2}.
The rules that govern how the agents interact, as well as some relevant parameters, are inferred from the data collected from an empirical chat system \texttt{Ubuntu} channel. (Note that, in this empirical system, a Bot is also present, but it is not emotional.)
In the model, the agents, representing users, exchange emotional messages with each other  and with the Bot. The chatting process thus depends on the dynamics of emotions (arousal and valence) of each agent, which fluctuate in time according to the impact of other agent's messages and the Bot's messages. In addition, other agent's attributes may affect their activity \cite{BT_ABMbook}. Specifically, each agent has a number of attributes as follows
\begin{equation}
A[i; (a_i(t),v_i(t)); profile; social.connections; delay; type; status] \ .
\label{eq-agent}
\end{equation}
Here, the agent's identity is fixed by an integer $i$ while its status---active or passive---is dynamically changing, depending on the events in its neighborhood on the network, its emotional state (high arousal triggers an action) and its activity pattern, taken by the delay time since the agent's $i$ previous action.
The dynamic variables $a_i(t)$ and $v_i(t)$ stand for the emotional arousal and valence components of an agent $i$, where  $i=1,2 \cdots N(t)$ and $N(t)$ is the number of agents in the network at time $t$. The network evolves by adding $p(t)$ agents per time step.  The agent's $i$ profile is fixed by its first appearance in the network; specifically, we fix the maximum number of messages $N_c^i$ that the agent $i$ can post during the simulation time  and the agent's attitude towards direct communications with other agents $g_i$ versus ($1-g_i$) posting messages on the channel, where they can be seen by all currently active agents. These parameters are taken from the empirical distributions  $g\in P(g)$ and $N_c^i\in P(N_c)$, which are inferred from the \texttt{Ubuntu} chats. The delay time, $\Delta t$ between consecutive actions of each agent is taken from the distribution $\Delta t \in P(\Delta t)$, which is also computed from the same empirical data. Motivated by the rules of the empirical system, in the model we also define  a certain number of moderators, whose activity pattern is slightly different from the ordinary agents. Namely, their delay-time distribution is fasted decaying function, also inferred from the empirical system. Moreover, similar as the Bot, the moderator agents have an unlimited number of messages $N_c^i=\infty$. Compared to other agents, the moderators have an additional function to fetch the messages posted on the channel (usually by newcomers) and reply to them or forward (the fraction $\epsilon$) to the Bot, then the Bot replies directly to the involved agent.
The dynamics of emotions is captured by two nonlinear maps for each agent, which are under the influence of temporally varying fields as described below and in refs.\ \cite{we-ABM_bots2,we-ABMrobots}. Specifically, the maps in the following form and polynomial nonlinearity can be used \cite{BT_ABMbook}
\begin{equation}
x_{i}(t+1)=
  (1-\gamma_{x})x_i(t) + \delta(\Delta t_i)\frac{h^{x}_{i}(t)+qh^{x}_{mf}(t)}{1+q}(1+c_{2}x_{i}(t)(1-x_{i}(t)^{k}))(1-|x_{i}(t)|).
   \label{valence_map}
\end{equation}
where, $x_i(t)$ stands for either arousal $a_i(t)$ or valence $v_i(t)$ of an agent $i$ at time $t$ and the index $i=1,2,\ldots,N_{a}(t)$, where $N_{a}(t)$ is the number of agnets at time $t$. 
To ensure the existence of a nontrivial fixed point, the nonlinear terms are added. In the case of arousal, the quadratic nonlinearity ($k=2$) is sufficient; on the other hand,  $k=3$ is necessary for the  valence map in order to have a fixed point both in the negative and the positive region.
The influence fields $h^{a}_{i}(t),  h^{v}_{i}(t)$ are computed from the messages directed to the agent $i$. On the other side,  the messages posted on the channel also contribute to the mean fields  $h^{a}_{mf}(t),  h^{v}_{mf}(t)$, which are perceived by all currently active agents. In particular, 
\begin{equation}
h^{a}_{i}(t)=\frac{\sum_{j\in
    {\cal{L}}_{in,i}}a^{m}_{j}(\theta(t_{m}-(t-1))-\theta(t_{m}-(t-T_{0})))}{\sum_{j\in   {\cal{L}}_{in,i}}(\theta(t_{m}-(t-1))-\theta(t_{m}-(t-T_{0})))} \ , 
\label{arousal_field}
\end{equation}
where $\theta(t)$ is Heaviside theta function,  $a^{m}_{j}$ is the arousal of the message arriving along the link $j$, and the summation is over the messages from the list  ${\cal{L}}_{in,i}$ of agent's $i$ incoming links; 
\begin{equation}
  h^{v}_{i}(t)=\frac{1-0.4r_{i}(t)}{1.4}\frac{N^{p}_{i}(t)}{N^{emo}_{i}(t)}-\frac{1+0.4r_{i}(t)}{1.4}\frac{N^{n}_{i}(t)}{N^{emo}_{i}(t)} \ .
\label{valence_field}
\end{equation}
Here,  $N^{p}_{i}$ and $N^{n}_{i}$ are the number of the messages within the considered time window which are directed to $i$  and  convey positive and negative emotion valence, respectively, and $N^{emo}_{i}(t)=N^{p}_{i}(t)+N^{n}_{i}(t)$. $r_{i}(t)=sgn(v_{i}(t))$ represents the sign of emotional valence of the agent $i$ at time $t$, which can affect the way the agent perceives a new message with the emotional content.
The corresponding components of the common fields  $h^{a}_{mf}$ and $h^{v}_{mf}$ are computed in a similar way, but considering the list of all recent messages on the channel, $\cal{S}$, including those that are not addressed to any particular agent:
The relaxation with rate $\gamma_a=\gamma_v$ is executed systematically. Whereas, the contribution from the nonlinear terms is  added according to the rules, i.e., when the agent's interactivity time $\Delta t$ expires.
According to the rules, an elevated  arousal may trigger agent's activity, thus both current arousal and valence of the agent are transmitted with the agent's message.

In contrast to references \cite{we-ABM_bots2,we-ABMrobots}, where the emotional Bots with a fixed emotion have been considered, the profiles of the Bots that we study here are the three chat Bots which are used in the experiment of  sect.\ \ref{sec-experiment}. In the model, these experimental Bots are implemented in the following way. First, we fix 91 agents (note that 40 of them are the moderators) with whom the Bot will exchange the emotional messages. The Bot has a table with $j=1,2 \cdots 91$ rows
\begin{equation}
 B_j =\{(a_j^1,v_j^1), (a_j^2,v_j^2), \cdots   (a_j^{35},v_j^{35})\}
\label{eq-sequence-j}
\end{equation}
where each pair $(a_j^k,v_j^k)$ indicates the $k$th message in the sequence directed to the agent $j$; the sequence is identical with the sequence that  the Bot used to chat with the \textit{ user $j$ in the experiment} of sect.\ \ref{sec-experiment}.  Then, in  communicates with the agent $j$, i.e., according to the rules of the model,  the Bot will use next message in the row $j$. When the sequence is exhausted, it starts again from the beginning. With all other agents the Bot communicates with messages which carry no emotional valence, i.e. $(a =0.5,v=0.0)$. A similar table is constructed for other Bots by using the corresponding experimental sequences of their messages.
Note that, in contrast to experiment,  the temporal pattern of the agent--Bot dialogs in the model are regulated by the events in the entire network of agents. That is, the involved 91 agents, as well as all other agents, will have a delay time to respond to the Bot; both, the agents and the Bot can be simultaneously involved in different dialogs. In this way, the Bot's emotional onto the selected 91 agents is competing with the impact of the social network in which they are embedded. On the other hand, through these agents' connections,  the  Bot's impact propagates over the entire network.
In the following, with the analysis of the simulated data for each of the three Bots separately, we first quantify the impact of these emotional Bots onto the network of agents. Furthermore, by comparing the patterns of activity of the selected 91 agents with the patterns of 91 isolated users in the experiment, we estimate the effects of social structure onto the emotional conduct of these agents.

\subsection{Spreading the Bots' impact over chat-network: the emergence of collective emotions\label{sec-spreading}}
\begin{figure}[h]
\centering
\begin{tabular}{cc}
\resizebox{20.8pc}{!}{\includegraphics{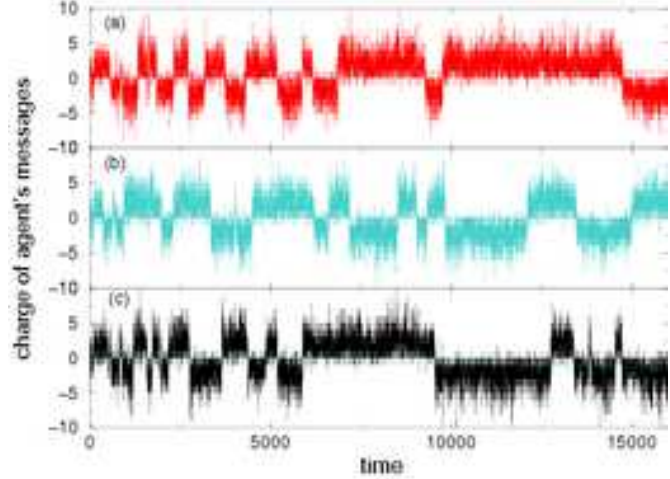}}\\
\end{tabular}
\caption{Charge of messages by all agents in the network (the Bot's messages are removed) in the presence of three experimental Bots with (a) positive, (b) neutral and (c) negative emotional profile. The straight lines indicate the increase/decrease of the average charge with time. 
}
\label{fig-abm-chargeExp}
\end{figure}

In contrast to experiment of sect.\ \ref{sec-experiment}, here the \textit{91 agents who receive the Bot's emotional messages} are interacting among each other and with the remaining agents in the evolving chat network, as shown in Fig.\ \ref{fig-network_exp_abm}.
Due to the interaction among agents, the emotional impact of Bots thus may spreads over the network. On the other hand, the emotional agents who are not under the direct influence of the Bot (i.e., not receiving the Bot's emotional messages) can influence these 91 agents through the direct links that they have established with them during the chatting process. In order to quantify the Bot's impact onto the entire network of agents, we compute the time series of emotional messages at the network level. Then constructing the  ``charge'' of these messages, $Q(t)=N^+(t)-N^-(t)$ we find out that the system of interacting agents slowly builds the emotional charge that coincides with the dominant emotion of the active Bot.
In Fig.\ \ref{fig-abm-chargeExp},  the time series of charge of emotional messages of the entire network is shown for the three  experimental Bots.  In all three cases, the charge fluctuations  indicate that initially the charge is balanced; then longer episodes of prevailing emotion start to occur, which coincide with the emotion of the active Bot. As a result, the positive charge builds up in the presence of the positive Bot while the negative charge starts prevailing in the case of the negative Bot.

\begin{figure}[!h]
\centering
\begin{tabular}{cc}
\resizebox{18.8pc}{!}{\includegraphics{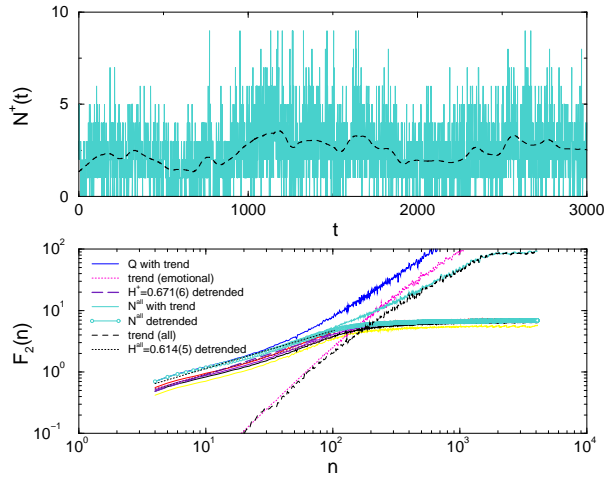}}\\
\end{tabular}
\caption{Top: Time series of all comment on the network in the presence of positive experimental  Bot sending emotional messages to 91 fixed agents and neutral messages to other agents. Dotted line indicates the trend of the time series. Bottom: Fluctuations of detrended time series of emotional messages in the presence of positive/negative Bot and all messages when the Bot is positive. For comparison, we also show how the trend affects fluctuations in the emotional charge (two top line) and in all messages (middle two lines).}
\label{fig-expBots-fluctuations}
\end{figure}
\begin{figure}[h]
\centering
\begin{tabular}{ccc}
\resizebox{16.8pc}{!}{\includegraphics{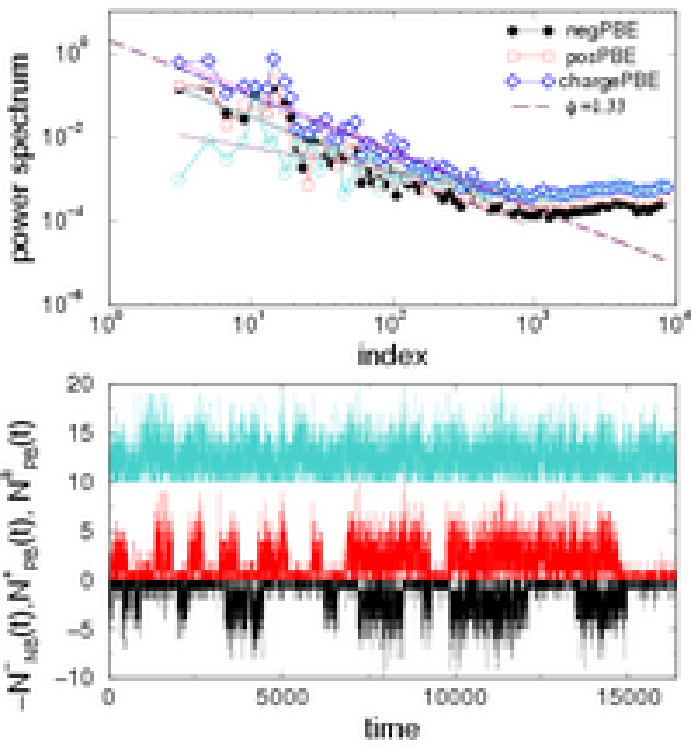}}&
\resizebox{16.8pc}{!}{\includegraphics{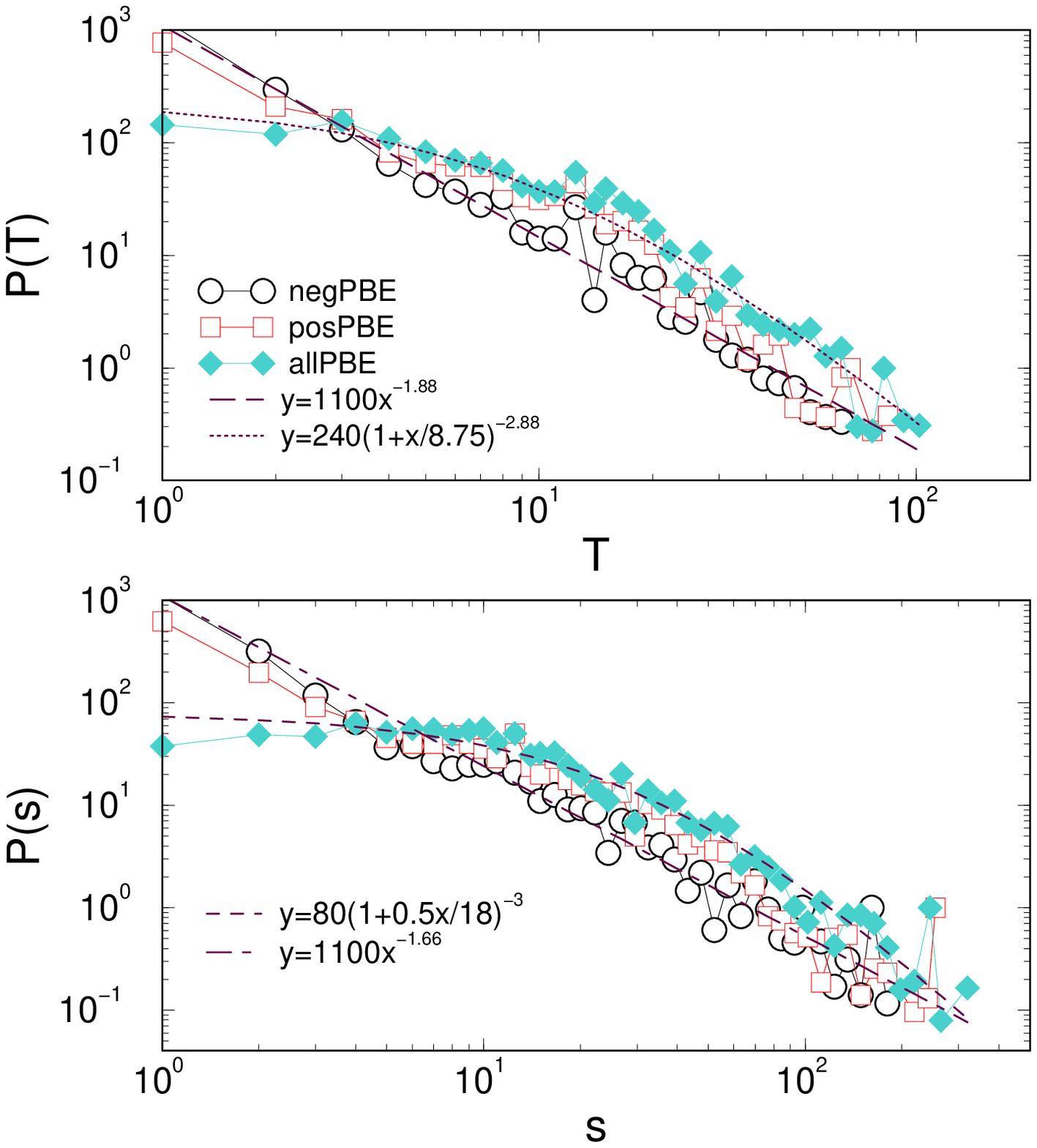}}\\
\end{tabular}
\caption{(bottom left) Time series of the number of negative messages in the presence of the negative Bot, $N^-_{NB}(t)$, and the number of positive messages, $N^+_{PB}(t)$ and all messages, $N^a_{PB}(t)$ (shifted vertically) in the presence of the positive Bot. The power spectrum of all time series and the charge for the positive Bot (top left).  The distribution of avalanche durations (top right) and sizes (bottom right) for the avalanches of all messages and messages carrying positive and negative emotion, as indicated, in the presence of the positive Bot. }
\label{fig-posBot-ts-avalanches}
\end{figure}

 In order to get further insight into the effects of the Bots, in section\ \ref{sec-compare} by examining the fluctuations of the emotional valence in the sequence of messages of all agents and the 91 agents under a direct influence of the Bot as well as grouping of these agents, we will estimate the impact of the network. 
Here, we first consider the fractal structure of the emotional time series at the level of the entire network.  In particular, the prominent characteristics of the underlying stochastic point process---temporal fluctuations, correlations and avalanches of the emotional messages---are altered when the Bots are active \cite{we-ABMrobots}. 
In the top panel of Fig.\ \ref{fig-expBots-fluctuations}, the time series of all messages of agents in the entire network is shown. These time series have a local cyclic trend (also shown by dashed line), which stems from the circadian cycles of the agents \cite{BT_ABMbook,we-ABMrobots}. By using the methods described in  \cite{we-MySpace11,we-ABMrobots} we remove the trend and analyze the fluctuations of the detrended time series around the trend. The standard deviations at the interval of the length $n$, $F_2(n)\sim n^H$, by which the Hurst exponent $H$ is defined,   are shown in the bottom panel of Fig.\ \ref{fig-expBots-fluctuations}. The corresponding Hurst exponent $H=0.614\pm 0.005$ indicates that we are dealing with a structured time series with    \textit{persistent} fluctuations. (Note that, for a random time series, the exponent $H=0.5$ is expected.) 
Performing a  similar analysis for the time series carrying positive messages in the presence of the positive Bot as well as for the negative messages in the presence of the negative Bot (these time series are depicted in Fig.\ \ref{fig-posBot-ts-avalanches}), we obtain the Hurst exponent in the same range. These fluctuations are also shown in  Fig.\ \ref{fig-expBots-fluctuations} (bottom panel) together with the fluctuations of the charge time series and the trend. Note that the scaling region (the area where the fluctuations obey a straight line) for the time series of charge is rather small, due to the tendency of the system towards balanced charge when the Bot is not active, that are further discussed in sect.\ \ref{sec-compare}.

The persistent fluctuations of the emotional time series suggest that further structure can be determined, i.e., in the temporal correlations and clustering of the events (avalanches), indicating the occurrence of \textit{collective behavior} of the agents.  Here, in Fig.\ \ref{fig-posBot-ts-avalanches} we show the results of such analysis of the time series in the presence of the positive Bot. Specifically, in the top left panel the power spectrum of the emotional time series and charge are shown to obey a power-law decay as $S(\nu) \sim \nu^{-\phi}$, with the scaling exponent $\phi\sim 1$ within error bars. The time series of all messages independently of their emotional content, however, is less correlated ($\phi =0.66$ is found).

The temporal clustering of events---avalanches can also be determined from these time series. An avalanche consists of the set of events that make the connected part of the time signal above a threshold line (a noise level) before it falls back to the threshold level \cite{tadic1996,spasojevic1996,dhar2006}; the  distance between two consecutive points where the signal meets the baseline determines the duration $T$ of an avalanche while the number of enclosed events is the avalanche size $s$.
In the right panels of Fig.\ \ref{fig-posBot-ts-avalanches} the distributions of the durations $P(T)$ and size $P(s)$ of the avalanches are shown. The avalanches are determined from the time series of all messages and the messages carrying positive and negative emotion valence as indicated in the legend.
As it can be expected in the presence of the positive Bot, the positive avalanches make a majority of clustered events. Moreover, the corresponding distributions $P(T)$ and $P(s)$ of the positive avalanches, as well as the avalanches of all messages, follow the $q$-exponential distribution \cite{we-ABMrobots} $P(X)=A(1+(1-q)X/X_0)^{1/1-q}$, with $q\approx 1.33$ within the numerical error bars. On the other hand, the avalanches of negative emotion messages exhibit  nearly a power-law decay with the exponent close to 1.66 and 1.88 for the size and duration distribution, respectively.

\subsection{Assessing the impact of social network\label{sec-compare}}
\begin{figure}[h]
\centering
\begin{tabular}{ccc}
\resizebox{18.8pc}{!}{\includegraphics{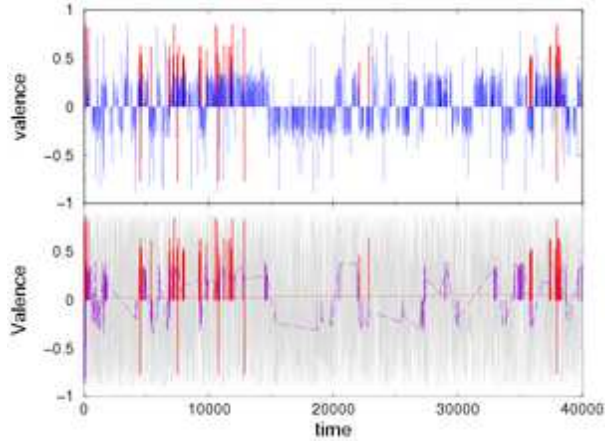}}\\
\end{tabular}
\caption{(top) Time series of valence in the messages of all agents communicated to the positive Bot (blue line). (bottom) Valence in the series of messages by all agents (gray line in the background) and of the 91 selected agents in the occasions when they communicated with each other (front line).
In both panels, the thick red line  indicates valences of the Bot's messages sent to the selected 91 agents. 
}
\label{fig-posBot-valence}
\end{figure}

The competition between the Bot's impact and the influence of the network onto the 91 agents who are in the direct contact with the Bot's emotional messages is examined. In Fig.\ \ref{fig-posBot-valence}, the valence fluctuation in the sequence of messages of these 91 agents are shown on top of the valence of all messages in the network and the positive Bot's messages. By its nature, the system of emotionally interacting agents has no preference towards either positive or negative emotions. 
However, when the Bot is active (and has a short interactivity time between consecutive messages), the number of positive messages among the selected 91 agents is larger than the number of negative messages. However, when the Bot's messages are sparse, both positive and negative messages occur, following the trend in the prevailing entire network. In the top panel, the valence of messages \textit{communicated to the Bot} from other agents in the network (note that the Bot's messages to them are neutral!) exhibits a similar fluctuation. That is, when the Bot's emotional activity is sparse, the oscillating valence sets in the emotional activity, which is also manifested in the messages communicated to the Bot.

\begin{figure}[h]
\centering
\begin{tabular}{ccc}
\resizebox{24.8pc}{!}{\includegraphics{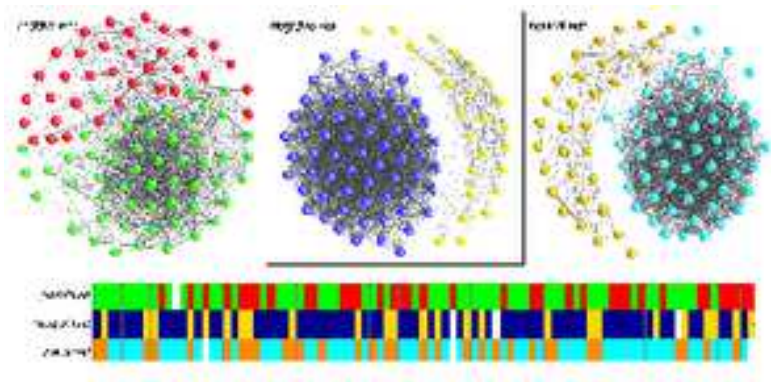}}\\
\end{tabular}
\caption{Communities induced by valence similarity in the 91 agents in the network, which are communicating with the emotional Bot with  positive, negative and neutral profile. In the lower figure, the color bars, e.g., green and red in the case of positive Bot, indicate the community to which the corresponding agent belongs; the agents $1,2, \cdots 91$ are ordered from left to right. }
\label{fig-communities_ABM}
\end{figure}

In analogy to the analysis of the experimental data in sect.\ \ref{sec-empirical}, here we  consider potential correlations among the specified 91 agents, who are in a direct contact with the emotional messages by the Bots.
The filtered correlation matrix with the widths of links above the applied threshold $W_0$ is mapped onto a graph. In Fig.\ \ref{fig-communities_ABM}, three graphs correspond to situations where the Bots with positive, negative and neutral emotion valence were  active. The community structure analysis led to  two communities that can be identified in each case--- a dominant (large)  and a subdominant (small)  community. Compared with the user groups in Fig.\ \ref{fig-groupsExp}, these communities are more compact. However, the identity of the nodes in each community vary when the Bot's emotion is changed.

\begin{figure}[h]
\centering
\begin{tabular}{ccc}
\resizebox{20.8pc}{!}{\includegraphics{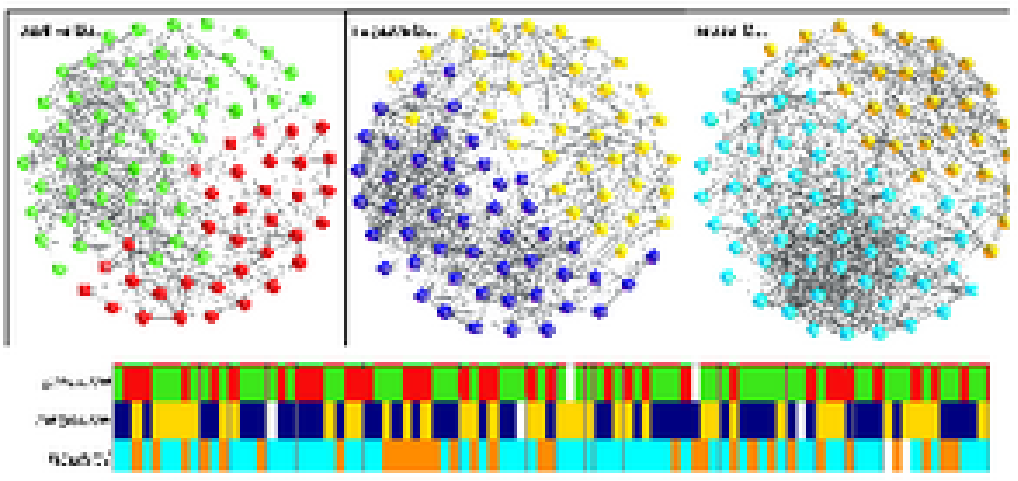}}&
\resizebox{14.8pc}{!}{\includegraphics{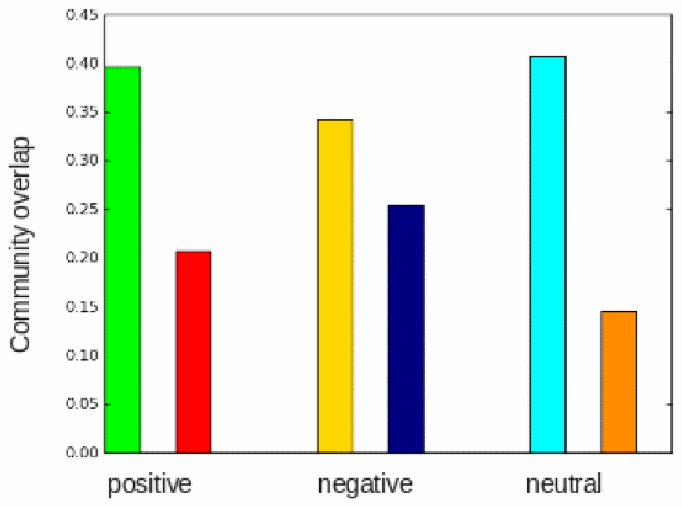}}\\
\end{tabular}
\caption{Left: Coarse grained user communities in analogy to Fig.\ \ref{fig-communities_ABM}. Right: Jaccard index for agent--user communities overlap for the Bots of different profiles.}
\label{fig-overlap}
\end{figure}

In order to assess the impact of the social network onto the formation of valence-based communities, we first coarse grain the community structure (i.e., using a reduced threshold in the community detection algorithm) in Fig.\ \ref{fig-groupsExp} so that we obtain only two communities---a smaller and a larger community, in analogy to communities of agents in Fig.\ \ref{fig-communities_ABM}. The resulting two-community structure is shown in Fig.\ \ref{fig-overlap} left. For comparison with  Fig.\ \ref{fig-communities_ABM}, the same color code is used; for example in the case of positive Bot, green or red color indicate that an agent belongs to the large or the small community, respectively.
Then, having the identity of each user (agent) in each community, we consider incidence that the user and the agent with the same index have the same color (i.e., belong to the same community). By other words, this gives the probability that a user, originally belonging to a community $C_k$ did not change the community under the influence of the network. The complementary events, however, measure the number of users who changed community due to the social network influence. 
Technically, we compute Jaccard index $J(A_k,U_k) = \frac{\vert A_k\cap U_k\vert}{\vert A_k\cup U_k\vert}$, where $A_k$, $k=1,2$ is a set of agents belonging  to the dominant and subdominant group, respectively. Similarly, $U_1$ and $U_2$ are sets of users in the respective large and small group.  For each emotional Bot, the values of $J(A_k,U_k)$  are shown in Fig.\ \ref{fig-overlap}(right). 
 Interestingly, the frequency of matching user--agent pairs varies when the Bot's polarity is changed.  By other words,  certain fraction of users, which grouped together under the influence of the Bot in the experimental conditions, will stay together within the same community inside the social network. However, some other users will switch community due to the social influence.  According to Fig.\ \ref{fig-overlap}, the largest dissimilarity between the agents and users groups, $d_k=1-J(A_k,U_k)$, is found in the case when the negative Bot is present.

\section{Conclusions}
We have studied appearance of groups and their collective behaviors that can be induced by emotional interactions with conversational Bots which conveyed different affective profiles, i.e., positive, negative and neutral. Considered are both the experimental environment, containing isolated users, and a social network environment, simulated within a suitable agent-based model. In both situations,  the Bots with exactly same emotional profiles (the same sequence of emotional messages) are employed.

To quantify the collective behaviors, we have analyzed the underlying stochastic process of emotional chats simulated with the agent-based model; for this purpose, the physics theory of complex systems is  used in accordance with graph theoretical methods. In addition, for analysis of the empirical data the algorithms for text analysis, which can extract emotional contents from communicated text messages, were employed.  The hallmarks of the chat process at a large scale are the self-organization, persistence and temporal clustering of emotional events. The present analysis suggests (see also \cite{we-ABMrobots}) that namely these features of the emotional chat processes enable the Bots to spread their influence on the entire network of agents, while communicating with a limited number of agents. The power spectrum, Hurst exponents and statistics of avalanches of emotional messages that are computed in section\ \ref{sec-spreading}, provide the corresponding quantitative measures of the collective behaviors of agents  in the presence of the Bots.
Apart from quantitative differences, the temporal correlations, as well as  the persistence of the process, are similar for Bots of all emotional profiles. However, the valence of the Bot's messages and messages of the agents communicated to the Bot are demonstrated to play a key role in inducing the overall charge of emotional messages in the entire network. In this respect, the positive Bot is a bit more effective than the negative Bot. Consequently, at the statistical level, the excess positive charge leads to increased probability of large positive avalanches. 

On the other hand, the participation of each individual agent (user) in the chat process and in making the network connections is different when the Bot's profile is changed. Comparing the experimental and simulated data, our main findings are summarized here:

\begin{itemize}
\item The Bots with positive/negative  profiles elicit the corresponding emotional reactions by users (and agents); the analysis of users' (and agents')  messages directed to the Bot reveals that their valence largely coincide with the Bot's profile. In addition, such conclusion is confirmed directly by users  in the related questionnaire, as a part of the experiment.
\item The score of positive emotion words in the presence of the positive Bot is higher than the score of negative emotion words in the case when the negative Bot is active. As possible reasons, we notice that, in the experiment, the users know that they are communicating with a machine. Hence, they assume rather “passive” stance; their emotion in the related responses might have been different in a setup where they would interact, e.g., with other users with such  characteristics. Further difference, taken into account by the model,  occurs in the scale of arousals of (often involved) positive and negative emotions  as well as their propagation in the social network.
\item The network theory analysis applied to the experimental data suggests that, even though the users are isolated in the chats with the emotional Bot, they can be grouped according to the similarity of their emotional reactions to the Bot's messages. Thus, these groups might originate from similarity of users' psychology profiles.  On the other hand, when exposed to the influence of a social network, which is simulated by the model, two coherent groups can be identified: a large community carrying the dominant emotion and the opposite emotion group. A fraction of  users that changed the group, when exposed to the social network influence, is larger in the case of  negative profile Bot than in the positive or neutral  Bot.
\end{itemize}

In conclusion, even though the social group activity  based on emotional interactions might be seen as  a primary subject of social psychology, it is necessary to use the interdisciplinary research  to deeper understand (and possibly control) a whole range of the related phenomena. The presented work shows  (see also \cite{we-Entropy} for the nonextensive emotion dynamics) that, in these multidisciplinary approaches,  the concepts of physics are indispensable. Specifically,  the physics methods reveal the  mechanisms in the underlying stochastic processes, through which the social groups emerge from individual actions of users. Furthermore,  within physics theory of complex systems,  the emergent collective emotional behaviors of the social groups can be quantified in the appropriate way.

\textbf{Authors' contribution:}  M.\v{S}. contributed software for simulations; M.S. provided experimental data and contributed sections 2.1 and 2.2; V.G. performed network analysis of experimental and simulated data and contributed related Figures; B.T. designed theoretical research, performed simulations and time-series analysis, contributed related Figures and wrote the paper.

\textbf{Acknowledgements:}
         We are grateful for support from program P1-0044 by
the Research Agency of the Republic of Slovenia and from the
European Community's program FP7-ICT-2008-3 under grant
no 231323.     B.T. also thanks for partial support from COST
Action TD1210 KNOWeSCAPE. M.S. would also like to
thank for support from projects OI171037 and III41011 of the
Republic of Serbia.


\end{document}